\documentclass[preprint,amsmath,amssymb]{revtex4}
\usepackage{graphicx}
\usepackage{tabularx}
\usepackage{makecell}
\usepackage{dcolumn}
\usepackage{bm}

\begin{document}
\title{Metamaterial analog of a black hole shadow 
: An exact ray-tracing simulation based on the spacetime index of refraction}
\author{${\rm M. \;Nouri}$-${\rm Zonoz}$ \footnote {Electronic
address:~nouri@ut.ac.ir\; (Corresponding author)}, ${\rm A.\;Parvizi}$ \footnote{Electronic
address:~a.parvizi@ut.ac.ir} and ${\rm H.\;Forghani}$\footnote{Electronic
address:~hasan.forghani@ut.ac.ir}}
\affiliation{Department of Physics, University of Tehran, North Karegar Ave., Tehran 14395-547, Iran.}

\begin{abstract}
In this letter first we show that the equation of null geodesics in spherically symmetric spacetimes in isotropic coordinates
is identical to the equation of light ray trajectories in isotropic media in flat spacetime. Based on this analogy we introduce an exact simulation 
of the light ray trajectories both in these spacetimes and in their metamaterial analogs in terms of the spacetime index of refraction.  
As unstable light trajectories, the photon spheres form in these metamaterial analogs at {\it exactly} the same radial distances 
as expected from the corresponding black hole geometries. Using the same ray-tracing simulation  we find the analog of a simple  black hole shadow formed by the 
metamaterial analog of a Schwarzschild black hole, eclipsing a line of light sources near its analog horizon. 
\end{abstract}
\maketitle 
%%%%%%%%%%%%%%%%%%%%%%%%%%%%%%%%%%%%%%%%%%%%%%%%%%%%%%%%%%%%%%
\section{Introduction} 
Metamaterial analogs of different spacetimes, based on the so called transformation optics \cite{Pen, Leon}, have attracted a lot of attention in recent 
years \footnote{Refer to \cite{Phil} and references therein.}.
Historically one could trace back this analogy to the analogy between a spacetime and a material medium  with respect to light propagation through which one could assign 
an index of refraction as well as other optical characteristics to the corresponding spacetime \cite{Edding,Gordon,LL,Pleban}. 
On the other hand the same analogy appears in the study of Maxwell's equations in a curved spacetime leading to constitutive equations with the geometric analog of 
the magneto-electric coupling (effect). 
Through this analogy, one could establish a correspondence between geometric entities of a curved spacetime and
the electromagnetic features of a  medium such as its electric permittivity and magnetic permeability \cite{LL,Pleban}.\\ 
In this opto-geometric relation one could start from a given spacetime and find its optical characteristics, and based on them design its optical analog using metamaterials 
in which the light trajectories mimic the null geodesics of the corresponding spacetime. Unlike the natural materials which have restricted  optical 
features, the metamaterial designs could demonstrate nontrivial optical features (such as negative index of refraction), and that is why one uses the term 
{\it metamaterial analogs} instead of optical analogs.
Obviously the more exotic a spacetime, the more interesting would be its optical features, and consequently their metamaterial analogs will enable one to realize
these exotic optical features. 
In other words these metamaterial analogs could help us to  examine interesting and perhaps observationally inaccessible 
optical characteristics of the corresponding spacetime. These could include, for example, interesting optical features associated with black hole spacetimes.
To this end one needs to simulate light rays, as exactly as possible, in the metamaterial analogs of the corresponding black hole spacetimes.
In previous studies full-wave simulations were employed to study light propagation in the metamaterial analogs of different spacetimes \cite{Chen,Fer,Ting} 
and then it was 
compared with the light ray trajectories in the geometric optics limit, formulated in the Hamiltonian language \cite{Fer} or through a ray-tracing mechanism \cite{Ting}.
Here we introduce a new direct and at the same time exact simulation of light ray trajectories in the metamaterial analogs of static spherically symmetric black holes, 
based only on their indices of refraction which are adapted from the corresponding spacetimes. 
The simulated trajectories are {\it exact} duplicates of those in the corresponding spacetime. In particular we find the analog of black hole photon spheres in their 
metamaterial analogs at the same exact radial distance as expected from the spacetime geometry. Indeed by increasing the precision level we could have rays orbiting 
the analog photon sphere as many times as the simulation cost allows.
This means that in the metamaterials designed with isotropic refractive indices borrowed from these black hole geometries, one could obtain photon 
spheres at specified radii. This could potentially find diverse applications both in electro-optical devices as well as in the investigation of the properties of so called 
{\it optical black holes} \cite{Unruh, Nariman}. One such interesting application is provided by showing how a simple metamaterial analog of a black hole shadow could be simulated.
%%%%%%%%%%%%%%%%%%%%%%%%%%%%%%%%%%%%%%%%%%%%
\section{Spacetime index of refraction}
Applying Fermat's principle to light rays in {\it stationary} spacetime, in the context of (1+3) or threading formulation of spacetime decomposition \cite{LL,DLBMNZ},
we are led to the following relation \cite{Besharat, NN}
\begin{equation}\label{dl}
\delta \int \left((\frac{1}{\sqrt{g_{00}}} + {\bf g} . {\bf {\hat k}}\right) dl_c = 0
\end{equation}
in which $\bf g$ is the so-called gravitomagnetic vector potential with components $g_{\alpha} = - \frac{g_{0\alpha}}{g_{00}}\;\; ( \alpha =1,2,3)$, ${\bf{\hat k}}$ is the unit 
vector along the ray, and $dl_c$ is the spatial line element in the curved spacetime \cite{LL,DLBMNZ}.
Obviously from the above equation one could assign the following index of refraction to the  3-space on which $dl_c$ is the spatial line element
between any two events in stationary spacetimes
\begin{equation}\label{in1}
n_s = n_0 + {\bf g} . {\bf{\hat k}}
\end{equation}
in which $n_0 = \frac{1}{\sqrt{g_{00}}}$ is the index of refraction assigned to a static spacetime 
($g_{\alpha} = 0$). It is noted that for a curved spacetime, the spatial line element is not necessarily flat.
Now if we restrict our attention to the static spacetimes and look for their 
(meta-)material analogs in a {\it flat} spacetime background, we could rewrite equation \eqref{dl} in the following general form
\begin{equation}\label{in2}
\delta \int \frac{1}{\sqrt{g_{00}}} (\frac{dl_c}{dl_f})dl_f = 0
\end{equation}
where $dl_f$ is the spatial line element in flat spacetime. The above relation shows that the metamaterial analog of a static spacetime could be 
assigned with the following index of refraction
\begin{equation}\label{in3}
n_f = \frac{1}{\sqrt{g_{00}}} (\frac{dl_c}{dl_f}).
\end{equation}
in which case the light ray  trajectories in the designed metamaterial mimic the null geodesics in the corresponding spacetime. The above argument shows that one 
could in principle assign the appropriate  index of refraction to the static spacetime if there is a coordinate system in which the spatial part of the metric is
conformally flat.\\
Specifically in the case of static spherically symmetric spacetimes with the general form of ($c=1$),
\begin{equation}\label{ds1}
ds^2 = d\tau^2 - dl^2_c = f(r)dt^2 - \left( \frac{1}{f(r)} dr^2 + r^2 d\Omega^2 \right),
\end{equation}
the line element could be transformed to the following {\it isotropic} from  by introducing the radial coordinate 
$\rho = const. \exp (\int \frac{1}{\sqrt{r^2 f(r)}}) $  \cite{Weinberg},
\begin{equation}\label{ds2}
ds^2 = f(r(\rho))dt^2 -F(\rho){dl^2_f}. 
\end{equation}
Using the above isotropic form of the spacetime metric and Eq. \eqref{in2}, the static spherically symmetric spacetimes, compared to the 
flat spacetime, is endowed with the following index of refraction \footnote{In the case of spherically symmetric spacetimes in isotropic coordinates 
one could see that $ds^2=0$ in \eqref{ds2} leads to $\frac{dl_f}{dt}= {\sqrt\frac{f(\rho)}{F\rho)}}=\frac{1}{n}$ \cite{Edding}.},
\begin{equation}\label{in4}
n_{sph} = {\sqrt\frac{F(\rho)}{f(\rho)}}
\end{equation}
For some spherically symmetric spacetimes $\rho (r)$ can be obtained 
analytically. These include the  Schwarzschild and  Reissner-Nordstrom  black hole geometries for which we are led to the following indices of refraction,
\begin{gather}
n_{Sch} = \frac{(1+\frac{M}{2\rho})^3}{(1-\frac{M}{2\rho})} \label{in51}  \\ 
n_{RN} = \frac{[(M+2\rho)^2 - Q^2]^2}{4 \rho^2 (Q^2-M^2+4\rho^2)} \label{in52}. 
\end{gather}
Since the isotropic coordinates only cover the exterior region of the black hole (e.g for Schwarzschild $2M < r < \infty$ or $M/2 < \rho < \infty$) \cite{Buch}, the above 
relations are also valid only for the {\it exterior region}, and so our simulation is naturally {\it limited to the same region} in the metamaterial analog.\\
%%%%%%%%%%%%%%%%%%%%%%%%%%%%%%%%%%%%%%%%%%%%%%%%%
\section{Null geodesics in spherically symmetric spcetimes in terms of the spacetime index of refraction}
Starting from the general form of the metric of  spherically symmetric spacetimes in isotropic  coordinates, namely \eqref{ds2}, we introduce the Lagrangian 
\begin{equation}\label{L1}
L = f(\rho)\dot{t}^2 - F(\rho)(\dot{\rho}^2 + \rho^2 \dot{\theta}^2 + \rho^2\sin{\theta}^2 \dot{\phi}^2 )
\end{equation}
where $^{.} \equiv d/d\lambda$ and $\lambda$ is an affine parameter along the null geodesics. Due to spherical symmetry, 
without loss of generality, we take geodesics on the equatorial plane $\theta = \pi/2$ for which $L$ reduces to
\begin{equation}\label{L2}
L = f(\rho)\dot{t}^2 - F(\rho)(\dot{\rho}^2 + \rho^2 \dot{\phi}^2 )
\end{equation}
Since the Lagrangian is independent of $t$ and $\phi$, we have the following two first integrals from the Euler-Lagrange equations,
\begin{equation}\label{L3}
f(\rho)\dot{t} = E
\end{equation}
\begin{equation}\label{L4}
F(\rho)\rho^2 \dot{\phi} = D
\end{equation}
representing the energy and angular momentum respectively.
Now since the spatial part of the spherically symmetric spacetimes \eqref{ds2} is conformally flat, the angles are given by the flat space formula, so for a 
small part of the ray trajectory making angle $\Theta$ with the radial coordinate we have (Fig.1),
\begin{equation}\label{ang1}
\sin{\Theta} = \frac{\rho d \phi}{\sqrt{\rho^2 d \phi^2 + d\rho^2}} = \frac{\rho \dot \phi}{\sqrt{\rho^2 {\dot\phi}^2 + {\dot\rho}^2}} 
= \frac{\rho}{\sqrt{\rho^2  + (\frac{d\rho}{d\phi})^2}}
\end{equation}
%%%%%%%%%%%%%%%%%%%%%%%%%%%%%%%%%%%%%%%%%%%%%%%%%%%%%%%%%%%%
\begin{figure}\label{0}
\includegraphics[scale=0.5]{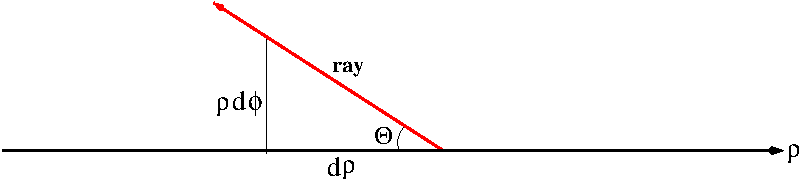}
\caption{Spatial geometry of a small portion of a light ray (in red) in spherically symmetric spacetimes in isotropic coordinates.}
\end{figure}
%%%%%%%%%%%%%%%%%%%%%%%%%%% %%%%%%%%%%%%%%%%%%%%%%%%%%%%%%%%%%%%%%%%%%%%%%%%%%%%%%%%
On the other hand for null rays $L=0$, and so from \eqref{L2} we have
\begin{equation}\label{L5}
f(\rho)\dot{t}^2 = F(\rho)(\dot{\rho}^2 + \rho^2 \dot{\phi}^2 )
\end{equation}
Now using the above equation along with equations \eqref{L3} and \eqref{L4}, to substitute for $\dot\rho$ and $\dot\phi$ in the second equality in \eqref{ang1}, we end up with
\begin{equation}\label{ang2}
\sin{\Theta} = \frac{1}{\rho}\frac{D}{E}\sqrt{\frac{f(\rho)}{F(\rho)}}=\frac{b}{\rho n(\rho)}
\end{equation}
in which we used  \eqref{in4}, and by definition $b=\frac{D}{E}$ is the impact parameter. Substituting $\sin{\Theta}$  back into the last equality in  \eqref{ang1},
we find the null geodesic equation in the following form
\begin{equation}\label{null}
 \frac{d\rho}{d \phi} = \rho \sqrt{\frac{\rho^2 n^2}{b^2}- 1}.
\end{equation}
This equation gives the null geodesic equation in terms of the spacetime index of refraction. \\
In what follows, we will be primarily concerned with the analog photon spheres and so we need to find the analog of the critical angle 
the rays should make with the 
radial direction (i.e angle of the cone of avoidance) to form the photon sphere. By the above considerations, and from equation \eqref{ang2} the critical 
angle at any given radial coordinate is found to be,
\begin{equation}\label{angle}
\sin \Theta_{cr} = \frac{b_{ph}}{\rho n(\rho)}
\end{equation}
where $b_{ph}$ is the impact parameter of rays forming the photon sphere at any radial coordinate $\rho$. 
In other words, the critical angle at a given radial coordinate is given in terms of the index of refraction 
at the same coordinate. Intuitively this is expected, as the 
combination $n (\rho) \sin \Theta_{cr} $ reminds one of the Snell's law, and the initial refraction needed at each radial coordinate for the rays to be trapped on 
the unstable photon sphere. 
%%%%%%%%%%%%%%%%%%%%%%%%%%%%%%%%%%%%%%%%%%%%%%%%%%%%%%
\section{Light ray trajectories in isotropic media}
The above mentioned analogy states that for metamaterial media designed with isotropic refractive indices given by \eqref{in51}-\eqref{in52}, 
the light trajectories would be the same as the null geodesics in the corresponding spacetimes. To verify this one should be able to 
simulate light rays in isotropic media. Interestingly enough in media with isotropic indices of refraction one could obtain 
the equation of light ray trajectories by considering their geometry in such media. This is already studied by Born and wolf in their classic text \cite{Born}. 
There it was shown that due to spherical symmetry, all the rays are plane curves satisfying the relation 
\begin{equation}\label{ray1}
 nr \sin \theta = C
\end{equation}
where $C$ is a constant and $\theta$ is the angle between the radius vector to a point on the light curve and the tangent to the curve at the same point (Fig. 2). 
To find the equation of light trajectories as plane curves, from the geometry in Fig. 2, it is noted that 
\begin{equation}\label{ray2}
\sin \theta = \frac{r(\phi)}{\sqrt{r^2(\phi) + (dr/d\phi)^2}}.
\end{equation}
Now substituting for $\sin \theta$ in the above equation  from \eqref{ray1},  we end up with the trajectory equation 
as \cite{Born},
\begin{equation}\label{ray3}
 \frac{dr}{d \phi} = r\sqrt{\frac{r^2 n^2}{C^2}- 1}.
\end{equation}
The similarity of equations  \eqref{ray1} and  \eqref{ang2} is a consequence of the fact that for spherically symmetric spacetimes, in isotropic coordinates, the spatial part 
of the metric is conformally flat. 
%Obviously the two radial coordinates are different in the two cases but that does not affect the ray simulations.
%%%%%%%%%%%%%%%%%%%%%%%%%%% %%%%%%%%%%%%%%%%%%%%%%%%%%%%%%%%%%%%%%%%%%%%%%%%%%%%%%%%
\begin{figure}\label{1}
\includegraphics[scale=0.35]{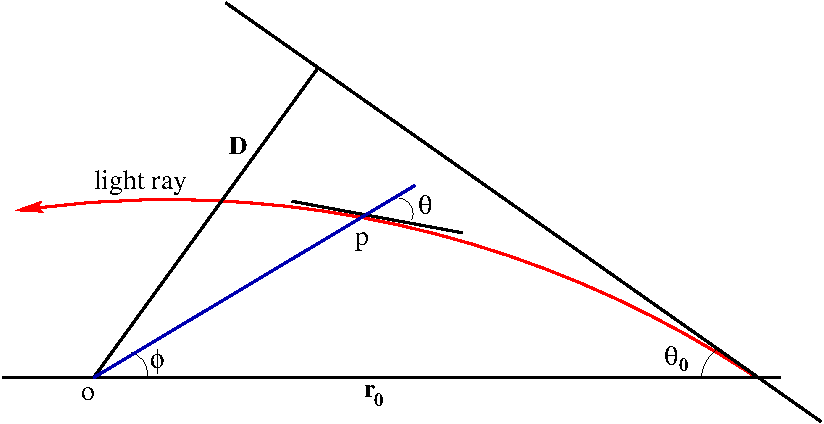}
\caption{Geometry of rays in an isotropic media.}
\end{figure}
%%%%%%%%%%%%%%%%%%%%%%%%%%%%%%%%%%%%%
\section{Simulation of light ray trajectories}
One can employ the equations \eqref{ray1} and \eqref{ray2} (or \eqref{ang2} and \eqref{ang3}) to simulate the light ray trajectories in an isotropic medium (or for that matter 
in a spherically symmetric spacetime in isotropic coordinates). 
For a light ray fired from $r=r_0$ along the $\theta_0$ direction, the impact parameter is $D=r_0 \sin \theta_0$ (Fig.2)
and, from \eqref{ray1}, $C = n(r_0) D$. Therefore the simulation goes as follows: for a given impact parameter and initial 
distance from the center of the metamaterial, the constant $C$ is fixed. To find the radial coordinate of the next point on the trajectory we substitute $\theta_0$ and $r_0$
in \eqref{ray2} to obtain the value of $\frac{dr}{d\phi}$. Having that, one can find the radial increment $\Delta r = \frac{dr}{d\phi} (\Delta \phi_0)$,
in which $(\Delta \phi_0)$ is the fixed-step increment of the azimuthal angle chosen according to the required precision. So now we have $r_1 = r_0 + \Delta r$, and one can 
repeat the same steps starting from \eqref{ray1}, now with $r=r_1$. The results of simulation for ray trajectories in different (isotropic) metamaterial analogs 
of spherically symmetric black holes are discussed next.
%%%%%%%%%%%%%%%%%%%%%%%%%%%%%%%%%%%%%%%%%%%%%%%%%%%%%
\subsection{Metamaterial analog of a Schwarzschild black hole}
The results of simulation for a congruence of light ray trajectories in a metamaterial analog of the Schwarzschild black hole 
leading to the formation of photon sphere are shown in Fig. 3. It is noted that the horizon, $r=2M $, 
and the photon sphere, $r=3M$  are mapped, in 
isotropic coordinates, to ${\rho}_h=M/2$, and $\rho_{ph}=(2 + \sqrt{3}) M/2$ respectively. The analog photon sphere 
is formed by 267 rays fired from $\rho = 36 (M/2)$ with the critical angle ranging from   $14.98166575^{\circ}$ to $14.98166577^{\circ}$ ($\delta\Theta_{cr} \sim 10^{-8}$).
In our simulations we  scanned a whole range of angles reaching the above value which matches exactly with
its theoretical value given by \eqref{angle} for Schwarzschild black hole, namely
\begin{equation}\label{cri1}
\sin \Theta_{cr} = \frac{3\sqrt{3}M}{\rho}\frac{(1-\frac{M}{2\rho})}{(1+\frac{M}{2\rho})^3} \equiv \frac{3\sqrt{3}M}{\rho n_{Sch}} 
\end{equation}
In this case the analog photon sphere is formed by rays with the closest distance of approach $\rho_{cda} \approx 3.7321 \rho_{\rm sch}$. 
All the rays rotate at least three times around the hole before escaping it.
To follow different rays we have included their blue-shift as they get closer to the analog photon sphere
with the help of a color gradient according to the medium's isotropic index of refraction. In the case of the metamaterial medium, both the speed and 
wavelength of the light ray change as it passes 
through the medium,  but its frequency remains intact. Obviously this optical feature is different from the {\it gravitational} blue shift as the light gets closer 
to the black hole horizon, in which case light frequency and wavelength change inversely but the speed of light is constant.
%%%%%%%%%%%%%%%%%%%%%%%%%%%%%%%%%%%%%%%%%%%%%%%%%%%%%%%%%%%%
\begin{figure}\label{2}
\includegraphics[scale=0.60]{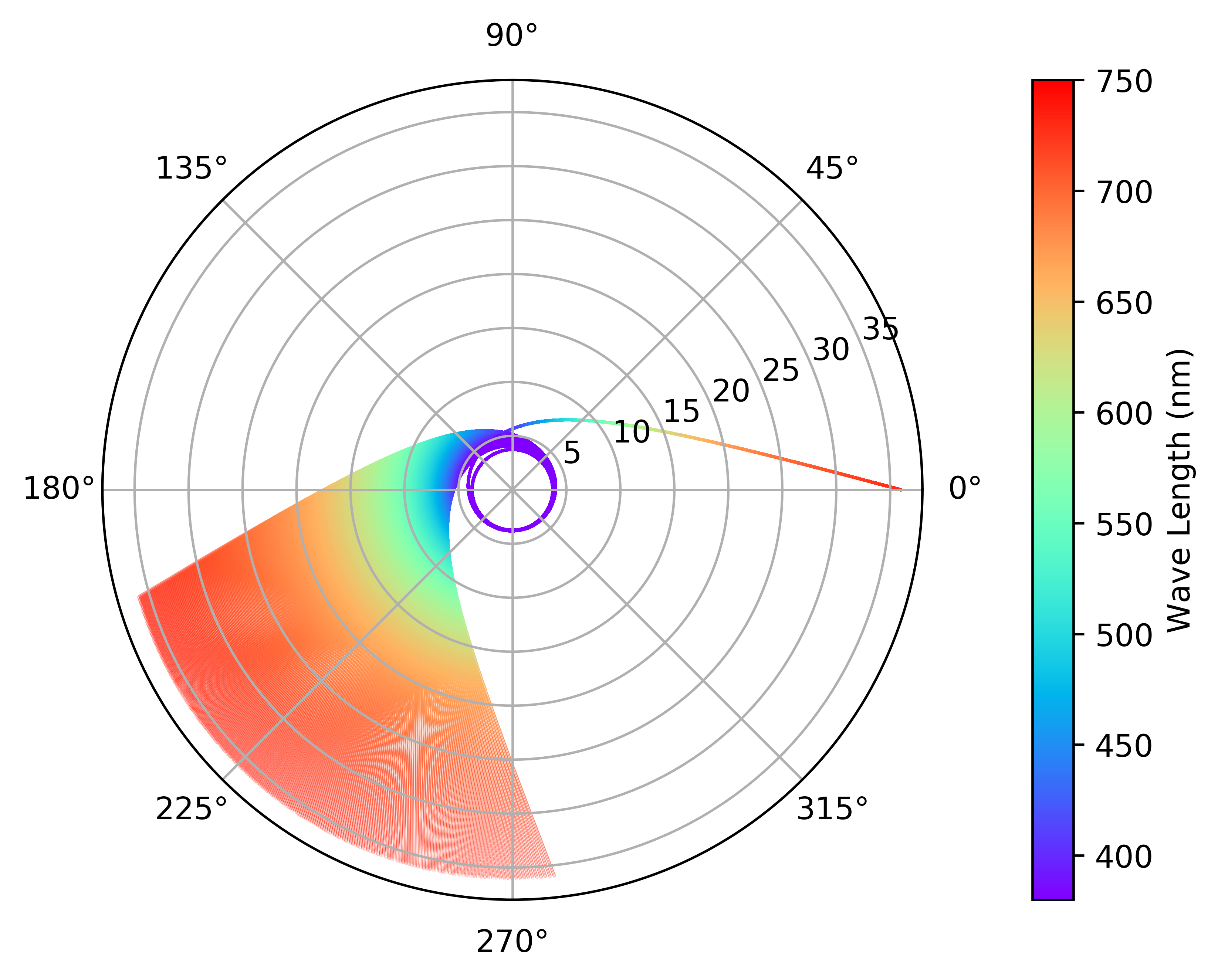}
\caption{Light ray trajectories in the metamaterial analog of a Schwarzschild black hole in isotropic coordinates and the formation of  
 photon sphere (purple circle). Distances are scaled to $M/2$.} 
\end{figure}
%%%%%%%%%%%%%%%%%%%%%%%%%%%%%%%%%%%%%%%%%%%%%%%%%%%
\begin{figure}\label{3}
\includegraphics[scale=0.60]{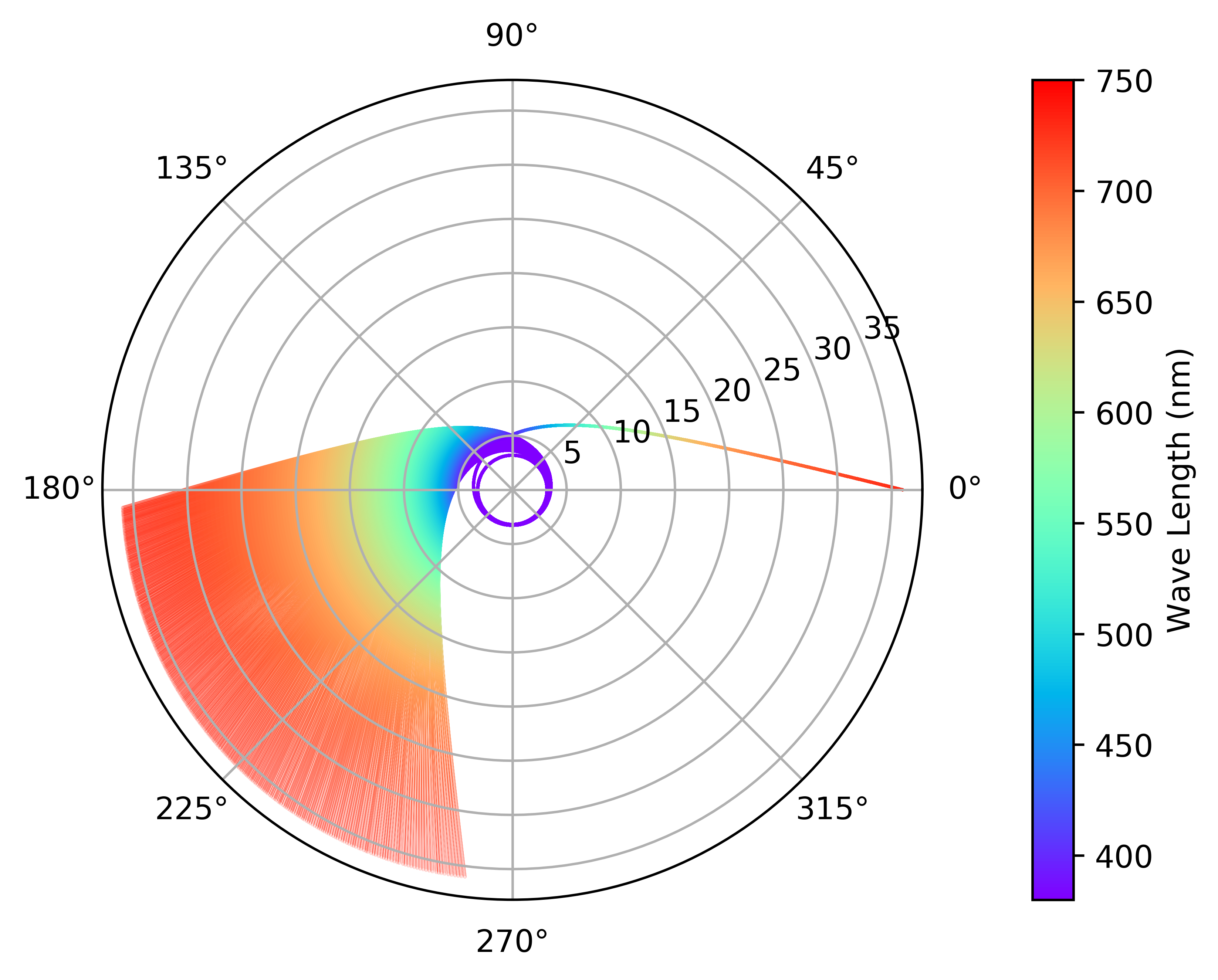}
\caption{Light ray trajectories in the metamaterial analog of Reissner-Nordstrom black hole with $Q/M=0.65$.}
\end{figure}
%%%%%%%%%%%%%%%%%%%%%%%%%%%%%%%%%%%%%%%%%%%%%%%%%%%%%%%%%5
\begin{figure}\label{4}
\includegraphics[scale=0.60]{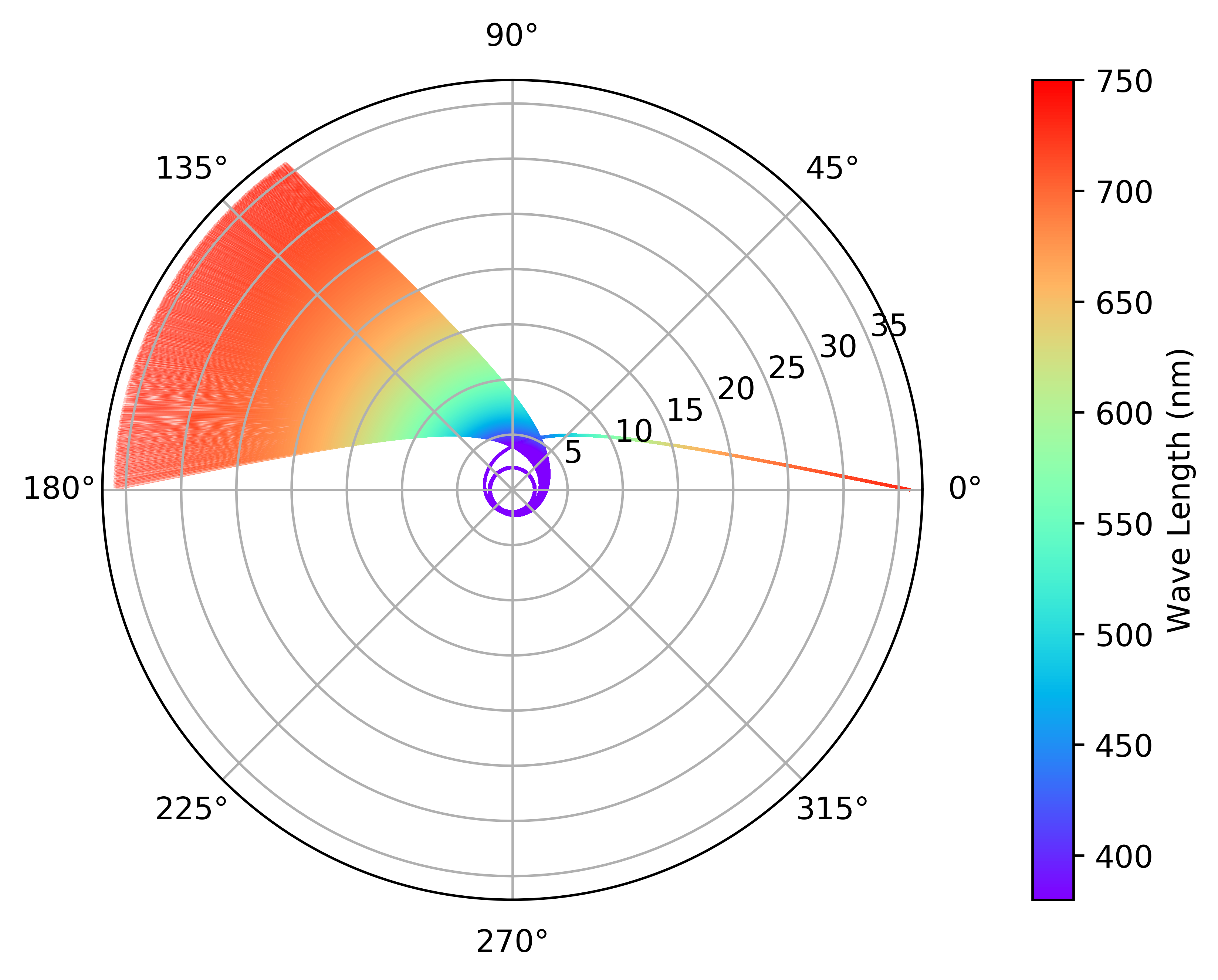}
\caption{Light trajectories in the metamaterial analog of an extreme Reissner-Nordstrom black hole, $Q/M=1$.}
\end{figure}
%%%%%%%%%%%%%%%%%%%%%%%%%%%%%%%%%%%%%%%%%%%%%%%%%%%%%%%%%%%%%%%%5
\subsection{Metamaterial analog of Reissner-Nordstrom black holes}
The results of  simulation for the ray trajectories in a metamaterial analog of Reissner-Nordstrom and extreme Reissner-Nordstrom (R-N) black holes with charges $Q=0.65 M$ 
and $Q=M$ are shown in Figures 4 and 5 respectively. 
The photon sphere is formed by two congruences of 334 and 351 rays fired at critical angles around $13.810519801^{\circ}$ and $11.50456352^{\circ}$ 
for $Q= 0.65M $ and $Q=M$ respectively. Again in complete agreement with their theoretical values given by \eqref{angle} for the corresponding indices of
refraction \eqref{in52}, and the impact parameters given by \cite{Chandra}
\begin{equation}\label{RN-imp}
b_{ph} = \frac{\frac{9}{4}M^2 [1+ (1-\frac{8Q^2}{9M^2})^{1/2}]^2}{\left( \frac{3}{2}M^2 [1+ (1-\frac{8Q^2}{9M^2})^{1/2}]- Q\right)^{1/2}}
\end{equation}
As expected from the R-N black hole geometry, compared to
the Schwarzschild black hole with the same mass, the analog photon spheres form at smaller radii in the corresponding metamaterial analogs.\\
%%%%%%%%%%%%%%%%%%%%%%%%%%%%%%%%%%%%%%%%%%%%%%%%%%%%%%%%%%%55
Obviously in the above simulations $M$ and $Q$ are just parameters in the metamaterial's isotropic index of refraction, Eqs. \eqref{in51}-\eqref{in52}.
Since the  metamaterial analog of R-N spacetime has two different parameters, one has more control on designing them with the required optical characteristics.
In our simulations the maximum winding number (number of rotations on photon sphere) was 5 for the extreme R-N case. Obviously we can reach higher winding numbers for each 
case by increasing computational precision and cost, but the present level of precision is high enough for our purpose.
Simulation parameters for the above three cases are listed in Table 1.
%%%%%%%%%%%%%%%%%%%%%%%%%%%%%%%%%%%%%%%%%%%%%%%%%%%%%%%%%%%%%%%%%%%%%%%%%%%%%%%
\begin{table}[h!]
\caption{Details of simulation for Metamaterial analogs of Schwarzschild and R-N black holes. $\Theta_{cr}$, $\rho_{\rm cda}$ and  $\lambda_m$  are the critical angle, 
the closest distance of approach and the minimum wavelength (maximum blue-shift) of the rays , respectively. All the rays have the initial wavelength of 730nm.
$\phi$ is azimuthal angle (in radian) representing the number of rotations around the hole (i.e the winding number).}
\label{table:1}
\centering
\begin{tabular}{p{2.5cm}p{2cm}p{3.5cm}p{2.5cm}p{2.5cm}p{1cm}c}
	\hline 
	\hline 
	$q = Q/M$& $\rho_{o}/\rho_{\rm sch}$ & $\Theta_{cr}$(in Radian)  & $\rho_{\rm cda}/\rho_{\rm sch}$ & $~~~~\phi$ & $\lambda_m$(nm) & \\ 
	\hline 
	0 & 36 & 0.2614793948 & 3.7321649 & $6\pi + 4.82$ & 292.75 &  \\ 

	0.65 & 36 & 0.2410390419 & 3.1896082 & $8\pi + 4.59$ & 270.94 &  \\ 

	1 & 36 & 0.2007925125 & 2.0000398 & $10\pi + 3.13$ & 203.34 &  \\ 
	\hline
	\hline 
\end{tabular}
\end{table}
%%%%%%%%%%%%%%%%%%%%%%%%%%%%%%%%%%%%%%%%%%%%%%%%%%%%%%%%%%%%%%%%%%%%%%%%%%%%%%%
\pagebreak
\section{Metamaterial analog of a black hole shadow}
As another interesting  application of the above exact simulation, we consider a  metamaterial analog of a simple Schwarzschild black hole shadow.
The results of simulation for a line of light sources placed at $5.07 \rho_{\rm sch} - 8.06 \rho_{\rm sch}$ and eclipsed by the analog of a black hole 
region are shown in Fig. 6. Light rays emanating 
from each source (as part of an analog accretion disk), within and at the edge of the corresponding cone of avoidance, are strongly 
deflected to reach a distant observer, forming the analog of a black hole shadow.
Those light rays within the cone of avoidances are lensed to form the outer ring, which is basically the analog of the Einstein ring \cite{Nar}. 
The inner ring is produced by those rays emanating at the edge of each cone of avoidance, rotating twice around the photon sphere before escaping to the observer positioned  
at $72\rho_{\rm sch}$. This means that, due to the spherical symmetry, an  observer/eye in that position will see the pattern shown in Fig. 7. The inner and outer 
rings  are formed by rays coming from 100 point sources placed on a line of length $ 2.99 \rho_{\rm sch}$. Widths of the outer and inner rings are 
$ 1.15 \rho_{\rm sch}$ and $ 3.75 \times 10^{-6} \rho_{\rm sch}$, and from the observer's position they are seen at subtended angles $ 0.898^{\circ}$ 
and $ 2.932^{\circ} \times 10^{-6}$ respectively. Color gradient in Fig.7 shows the density 
of rays in each ring, and as expected, the relativistic ring (inner ring) is denser and hence  brighter than the outer lensed ring (we have to modify the color
gradient to make the difference between the two rings visually clear).\\
Obviously experimental realization of all the above phenomena in metamaterials needs a very delicate 
design specifically noting that our simulation is only valid for the region outside the analog horizon in the designed metamaterial. So for example in the case of the above
analog black hole shadow one needs sources whose light rays are emanating only inside the cone of avoidance to avoid those falling toward the center.
%%%%%%%%%%%%%%%%%%%%%%%%%%%%%%%%%%%%%%%%%%%%%%%%%%%%%%%%%%%%%%%%
\begin{figure}\label{5}
\includegraphics[scale=0.70]{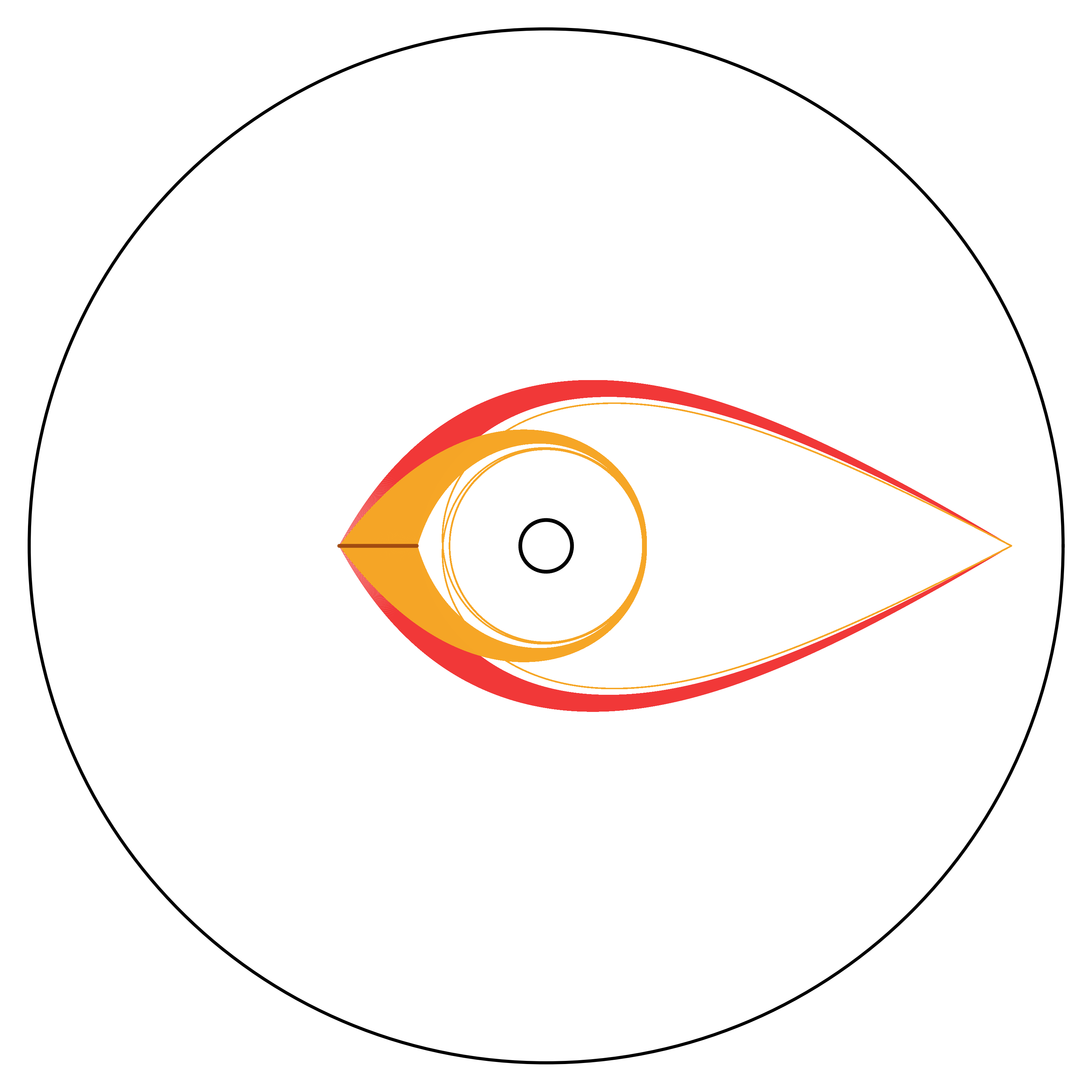}
\caption{The schematic side view of  metamaterial analog of a black hole 
	shadow produced by rays (in yellow ) from a line of light sources (the black line) behind its analog horizon (the black circle), reaching a distant 
	observer after rotating twice on its analog photon sphere (yellow circles). Other rays (in red) could reach 
	the observer directly with larger impact parameters.}
\end{figure}
%%%%%%%%%%%%%%%%%%%%%%%%%%%%%%%%%%%%%%%%%%%%%%%%%%%%%%%%%%%%%
\begin{figure}\label{6}
\includegraphics[scale=0.7]{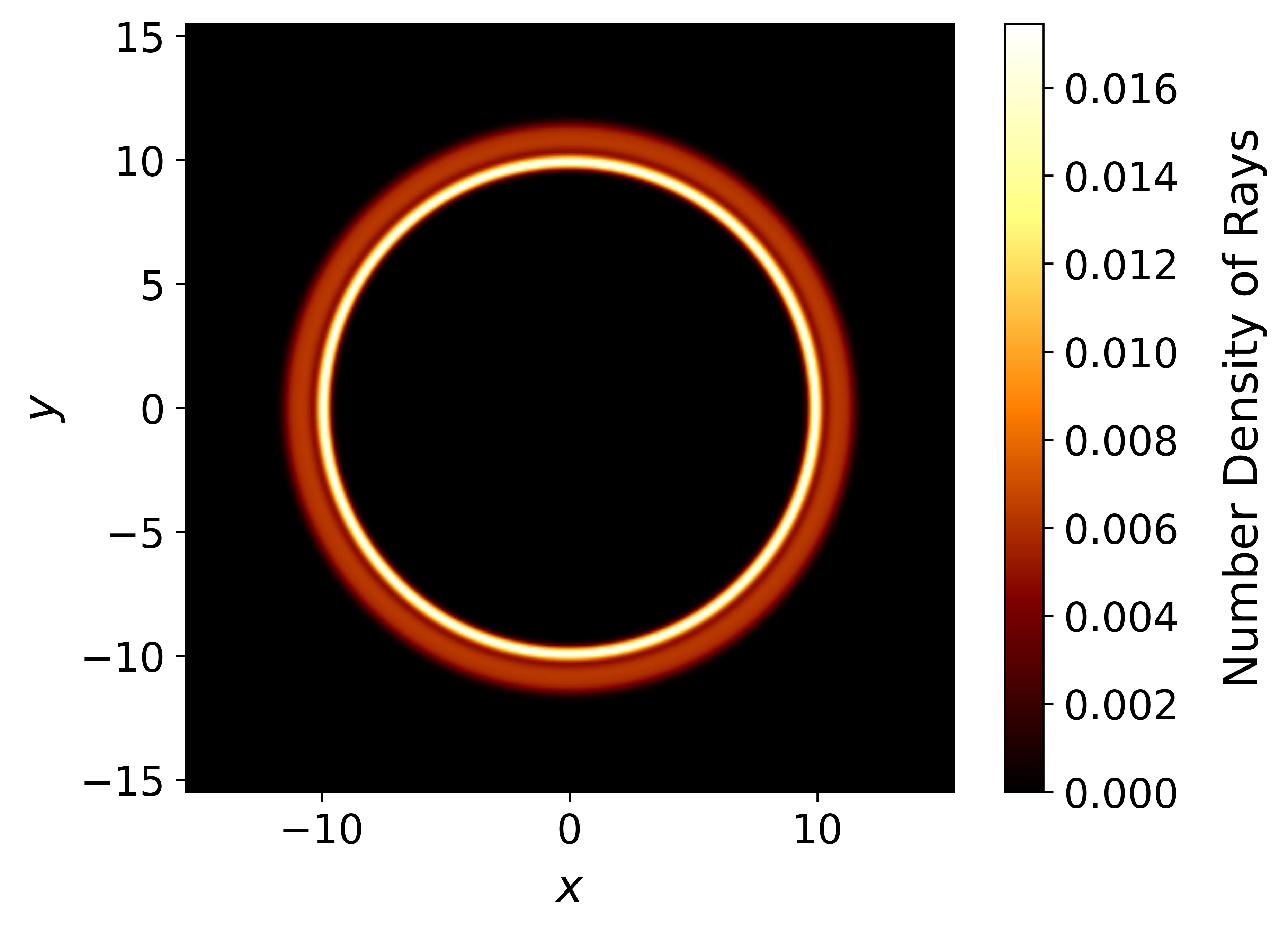}
\caption{The metamaterial analog of a black hole shadow as seen by an observer 
	at $\rho_{o} = 72 \rho_{\rm sch}$. It is produced by rotating the side view about the optical axis.}
\end{figure}
%%%%%%%%%%%%%%%%%%%%%%%%%%%%%%%%%%%%%%%%%%%%%%%
\section{Conclusions}
In this study we  investigated the metamaterial (optical) analog of spherically symmetric 
spacetimes based only on the metamaterial's index of refraction which was adapted from the spacetime geometry in isotropic coordinates. 
This was done by showing that the equation of null geodesics in spherically symmetric spacetimes (in isotropic coordinates)
is identical to the equation of light ray trajectories in media with isotropic indices of refraction.
Using this relation we introduced an exact ray tracing simulation both for the null trajectories (geodesics) in spherically symmetric spacetimes and 
light rays in isotropic metamaterial. Taking the metamaterial's refractive index to be identical to the spacetime index of refraction (modulated with its spatial conformal factor), 
it was shown that the structure of light ray trajectories in the  metamaterial exactly mimics that of the corresponding spacetime. This was explicitly shown for the case of analog  
photon spheres in the metamaterial analogs of Schwarzschild and R-N black holes .
The main advantage of the procedure outlined here is its simplicity and accuracy which allows for simulation of any spherically symmetric spacetime which could be written 
in isotropic coordinates. Finally the same simulation method was employed to find the optical analog of a simple black hole shadow formed by rays escaping the  photon sphere
of an analog of a Schwarzschild black hole.
%%%%%%%%%%%%%%%%%%%%%%%%%%%%%%%%%%%%%%%%%%%%%%%%%%%%5
\section *{Acknowledgments}
The authors would like to thank University of Tehran for supporting this project under the grants provided by the research council. They also thank 
the department of physics for using its HPC system, and I. Maleki for helping to run the simulations on the cluster. M.N-Z thanks Kh. Hasani for useful discussions. 
This work is based upon research funded by Iran national science foundation (INSF) under project No.4005058.
%%%%%%%%%%%%%%%%%%%%%%%%%%%%%%%%%%%%%%%%%%%%%%%%%%%%%%

\end{document}